\documentclass[aps,prb,showpacs,twocolumn]{revtex4}
\usepackage{amsmath}
\usepackage{amssymb}
\usepackage{epsfig}

\begin{document}
\title{Global Partial Density of States:
Statistics and Localization Length in Quasi-one Dimensional
disordered systems}
\author{J. Ruiz}
\affiliation{Departamento de F\'{\i}sica, Universidad de Murcia,
Apartado4021, E-30080 Murcia, Spain}
\author{E. J\'odar}
\affiliation{Universidad Polytechnic de Carageen, Departamento de
F\'{\i}sica Aplicada, Murcia, E-30202 Spain}
\author{V. Gasparian}
\affiliation{Department of Physics, California State University,
Bakersfield, CA, USA}

\begin{abstract}
We study the distributions functions for global partial density of
states (GPDOS) in quasi-one-dimensional (Q1D) disordered wires as
a function of disorder parameter from metal to insulator. We
consider two different models for disordered Q1D wire:  a set of
two dimensional $\delta$ potentials with an arbitrary signs and
strengths placed randomly, and a tight-binding Hamiltonian with
several modes and on-site disorder. The Green functions (GF) for
two models were calculated analytically and it was shown that the
poles of GF can be presented as determinant of the rank $N\times
N$, where $N$ is the number of scatters. We show that the
variances of partial GPDOS in the metal to insulator crossover
regime are crossing. The critical value of disorder $w_c$ where we
have crossover can be used for calculation a localization length
in Q1D systems.
\end{abstract}
 \maketitle

\section{Introduction}
\label{sec:introduction}

Calculation of density of states (DOS) allowed us obtain many
properties of the system under consideration, such as charging
effects, electrical conduction phenomena, tunneling spectroscopy
or thermodynamic properties. Furthermore, the decomposition of DOS
in partial density of states (PDOS) and global PDOS (GPDOS), which
appear naturally in scattering problems in which one is concerned
with the response of the system to small perturbation $\delta
U(x)$ of the potential $U(x)$, plays an important role in dynamic
and nonlinear transport in mesoscopic conductors
\cite{Buttiker93,Buttikeral94,Brand98,Zhengal97,Gasparian96,Buttiker01,Schomerusal02,Buttiker02}.
Particularly the emissivity, which is the PDOS in configuration
space for electrons emitted through arbitrary lead
\cite{Buttikeral94,B98,SM99}, always present in physical phenomena
where quantum interference is important.  As shown in \cite{MB02}
the heat flow, the noise properties of an adiabatic quantum pump
can expressed in terms of generalized parametric emissivity matrix
$\nu[X]$ (the diagonal element $\nu_{\alpha\alpha}[X]$ of which is
the number of electrons entering or leaving the device in response
to small change $\delta U(x)$, such as a distortion of the
confining potential). The nondiagonal element
$\nu_{\alpha\beta}[X]$ of parametric emissivity matrix determines
the correlation between current in the contacts $\alpha$ and
$\beta$ due to a variation of parameter $X$ \cite{MB02}. Note that
the elements of GPDOS are closely related to characteristic times
of the scattering process, consequently, to the absolute square of
the scattering states. Particularly in 1D systems
$\nu_{\alpha\beta}$ and $\nu_{\alpha\alpha}$ are related to Larmor
transmitted time $\tau_T$ (or Wigner delay time) and reflected
time $\tau_R$ weighted by the transmission coefficient $T$
\cite{Gasparian96,Prigodin99} and reflection coefficient $R$,
respectively. As it was mentioned by B\"uttiker and Christen
\cite{BC98} the dynamic response of the system to an external time
dependent perturbation, i.e., the emittance in general not
capacitance-like, i.e. the diagonal and the off-diagonal emittance
elements are not positive and negative values, respectively.
Whenever the transmission of carriers between two contacts
predominates the reflection, the associated emittance element
changes sing and behaves inductance-like. This type of cross over
behavior for diagonal element of emittance $\nu_{\alpha\alpha}$
(taking into account the Coulomb interaction of electrons inside
the sample) was found in \cite{Tiago00} where they study the
distribution function (DF) of emittance. They have found that in
the range of weak disorder, when the system is still conducive the
DF is Gaussian-like. With increasing disorder the DF becomes
non-Gaussian.

The purpose of this paper is to study numerically the behaviors of
DF of diagonal $\nu_{\alpha\alpha}$ and off-diagonal elements
$\nu_{\alpha\beta}$ of global PDOS in the Q1D disordered wires,
where not so much known about the DF. We study three different
regimes of transport: metallic ($\xi>>L$), where $\xi$ is the
localization length and $L$ the typical size of the system,
insulating ($\xi<<L$) and crossover ($\xi\sim L$). We show that in
intermediate regime of transport between the metallic and
insulating regimes there is the critical value of disorder $w_c$
when we observe cross over between the variances
var($\nu_{\alpha\alpha}$) and var($\nu_{\alpha\beta}$) (see
Fig.1). This critical $w_c$ determines the localization length of
Q1D system for given length $L$ and number of modes $M$. It turns
out that in metallic regime $P(\nu_{\alpha\beta})$ is Gaussian
which means that the first and the second moments (i.e., the
average $<\nu_{\alpha\beta}>$ and the variance
var($\nu_{\alpha\beta})=<\nu_{\alpha\beta}^2>- {<
\nu_{\alpha\beta}>}^2$) are enough to describe the behavior of
$P(\nu_{\alpha\beta})$. In the strong localization regime the
distribution of $\nu_{\alpha\beta}$ is log normal, which means
that the $\ln \nu_{\alpha\beta}$ follows a Gaussian distribution.
As regards the distribution function of $\nu_{\alpha\alpha}$ we
can say that in the strong localization regime it characterized by
an exponential tail, the values of $\nu_{\alpha\alpha}$ are
positive and that the dynamic response of the system is
capacitivelike\cite{BC98}. In the metallic regime the emittance
has non Gaussian-like behavior and some of the value of
$\nu_{\alpha\alpha}$ are negative (inductivelike
behavior)\cite{Tiago00}.

The paper is organized as follows. In the next section we present
our model and set the basis on numerical calculation for obtaining
the probability distributions of $\nu_{\alpha\alpha}$ and
$\nu_{\alpha\beta}$ for different regimes. In section
\ref{sec:means} we study the behavior of var($\nu_{\alpha\alpha}$)
and var($\nu_{\alpha\beta}$) as a function of disorder strength
$w$. In section \ref{sec:plots} we calculate the distribution
functions for $\nu_{\alpha\alpha}$ and $\nu_{\alpha\beta}$ in
three different regimes of transport mentioned in Introduction.
The paper is included in section \ref{sec:conclusions}.

\section{The model and numerical procedures}
\label{sec:model} The localization length $\xi$ is obtained from
the decay of the average of the logarithm of the conductance, $\ln
g$, as a function of the system size $L$
\begin{equation}
\xi^{-1}=-\lim_{L\rightarrow\infty}\frac{1}{2L}<\ln g>
\label{eq:green}
\end{equation}
where $g$ is given by B\"uttiker-Landauer formula
\cite{Landauer89,Datta95} ($T_{nm}$ is the transmission
coefficient from mode $n$ to mode $m$)
\begin{equation}
g=\frac{2e^2}{h}\sum_{n,m}T_{nm}, \label{eq:green2}
\end{equation}
We will consider two models: Q1D wire with the set of $\delta$
scattering potentials of the form
\begin{equation}
V(x,y)=\sum_{n=1}^{N}V_{n}\delta (x-x_{n})\delta (y-y_{n}),
\label{delta}
\end{equation}
with $V_n$, $x_n$ and $y_n$ be arbitrary parameters and Q1D
lattice of size $L\times W$ ($L\gg W$, $L$ is the length and $W$
is the width of the system), where the site energy can be chosen
randomly. In both cases analytically we have calculated the
Green's function of 1QD (\cite{RG06}) and use them in our
numerical calculations (see Appendix). The elements of global PDOS
$\nu_{\alpha\beta}$, in the case of a tight-binding model can be
calculate in terms of the scattering matrix and the Green
Function. To calculate the scattering matrix elements,
corresponding to transmission between modes {\sl n} and {\sl m},
we start from the Fisher-Lee relation \cite{Fisherlee81,Datta95},
which expresses these elements in terms of the Green's function:
\begin{equation}
s _{nm}={-\delta _{nm}+\imath\hbar\sqrt{v _n v _m}\sum_{i,j}\chi
_n\left(r_{0 _{i}}\right)G\left(r_{0 _{i}},r_{0_{j}}\right)\chi
_m\left(r _{0_{j}}\right)}. \label{eq:matrizsdef}
\end{equation}
$\chi _m\left(r _{0_{j}}\right)$ is the transverse wavefunction
corresponding to mode {\sl m} at the point $r _{0_{j}}$ and
$G\left(r _{0_{i}},r _{0_{j}}\right)$ is the Green's function (GF)
for non coinciding coordinates. $v_m$ is the velocity associated
with propagating mode $m$. The LPODS is directly connected to the
S-matrix elements $s_{nm}$ through the expression
\cite{Buttiker93}:
\begin{equation}
\frac{dn_{nm}(r)}{dE}\equiv-\frac{1}{4\pi\imath} \left(s_{nm}^*
\frac{\delta s_{nm}}{\delta U(r)}-\frac{\delta s_{nm}^*} {\delta
U(r)}s_{nm}\right) \label{eq:matrizndef}
\end{equation}
Insertion of Eq. (\ref{eq:matrizsdef}) in Eq.
(\ref{eq:matrizndef}) gives:
\begin{widetext}
\begin{equation}
\frac{dn_{nm}}{dE}(r)=- \frac {\hbar \sqrt{v _n v _m}}{4\pi }
\sum_{i,j}\left(s_{mn}^*\chi_n(r_{0_{i}})G(r_{0_{i}},r)G(r,r_{0_{j}})\chi_m(r_{0_{j}})\break
+H.c.\right) \label{eq:matriznapl}
\end{equation}
\end{widetext}
where H.c. denotes Hermitian conjugate. To arrive the above
expression we have calculated the functional derivative of the
Green's function by adding to the Hamiltonian of our system the
local potential variation $\delta U(r)=\delta U_a\delta (r-r_a)$
($\delta U_a\rightarrow 0$), which lead us to the relation
\cite{Gasparian96}
$$\frac{\delta G(r_n,r_m)}{\delta U(r)}=G(r_n,r)G(r,r_m).$$
Once we have calculated the local PDOS we can obtain the global
PDOS adding the local PDOS over the particles of our system:
\begin{equation}
\frac{dN_{nm}}{dE}=\sum _k \frac{dn_{nm}(r_k)}{dE}
\label{localglobal}
\end{equation}
After summation over the indices $i,j$ and $r_k$ the above
equation in matrix form can be presented:
\begin{equation}
\frac {dN_{nm}}{dE}=- \frac {\hbar \sqrt{v _n v _m}}{4\pi }
(s_{mn}^*Q_{nm}+H.c)
\end{equation}
where $Q_{nm}$ matrix defined as
\begin{equation}
Q_{mn}=\tilde{\chi}_n\left(\sum_{j=1}^MG_{mj}G_{jn}\right)\tilde{\chi}_m^T
\end{equation}
$\tilde{\chi_n}$ is the column matrix:
\begin{equation}
\tilde{\chi}=\left(
\begin{array}{c}
\chi_n(1)\\
\vdots\\
\chi_n(M)
\end{array}
\right)  \label{s5}
\end{equation}
Here $\tilde{\chi}_m^T$ is the transpose of the column matrix
$\tilde{\chi}_m$ and $G_{mj}$ is the matrix of $M\times M$ rank
($M$ is the number of modes in each lead, see Appendix).

Finally, $\nu_{\alpha\alpha}$ and $\nu_{\alpha\beta}$ one can get
from global PDOS ${dN_{nm}}/{dE}$ by summing every mode $n$ in
lead $\alpha$ and every mode $m$ in lead $\beta$ respectively:
\begin{equation}
\nu_{\alpha\alpha}=\sum_{n,m \in \alpha}\frac{dN_{nm}}{dE}
\label{pdr}
\end{equation}
\begin{equation}
\nu_{\alpha\beta}=\sum_{n\in \alpha,m\in \beta}\frac{dN_{nm}}{dE}
\label{pdt}
\end{equation}
Similarly can be written also $\nu_{\beta\beta}$ and
$\nu_{\beta\alpha}$ and so global DOS $\nu$ must be sum of all
GPDOS:
\begin{equation}
\nu=\nu_{\beta\beta}+\nu_{\alpha\alpha}+\nu_{\alpha\beta}+\nu_{\beta\alpha}.
\label{dos}
\end{equation}
In the case of the Q1D wire with the set of $\delta$ potentials
(see Eq. (\ref{delta})) in quantities (\ref{localglobal}),
(\ref{pdr}) and (\ref{pdt})), calculated for tight-binding model
one must replace the sign of summation by appropriate spatial
integration.

For numerical study we consider a quasi-one dimensional lattice of
size $L\times W$ ($L\gg W$), where $L$ is the length and $W$ is
the width of the system. The standard tight-binding Hamiltonian
with nearest-neighbor interaction
\begin{equation}
H=\sum_i \epsilon_i \mid r_i> <r_i\mid -t\sum_{i,j}\mid r_i><
r_j\mid, \label{eq:hamiltoniano}
\end{equation}
where $\epsilon_i$ is the energy of the site $i$ chosen randomly
between $\left(-\frac{w}{2},\frac{w}{2}\right)$ with uniform
probability. The double sum runs over nearest neighbors. The
hopping matrix element $t$ is taken equal to $1$, which sets the
energy scale, and the lattice constant equal to $1$, setting the
length scale. The energies are measured with respect to the center
of the band so we will always deal with propagating modes. Finally
our sample is connected to two semi-infinite, multi-modes leads to
the left (lead $\alpha$) and to the right (lead $\beta$). For
simplicity we take the numbers of modes in the left and right
leads to be the same ($M$) and so the width of this system $w$
becomes equal $M$ (for a tight-binding model the numbers of modes
coincides with the number of sites in the transverse direction).
The conductance of a finite size sample depends on the properties
of the system and also on the leads which must be taken into
account in a appropriate way. In order to take into account the
interaction of the conductor with the leads we introduce a
self-energy term "$A$" as an effective Hamiltonian, which will be
calculated as (see, e.g. \cite{Datta95})
\begin{equation}
A_p(r_{0_{i}},r_{0_{j}})=-t\sum_{m \epsilon
p}\chi_m(r_{0_{i}})e^{\imath k_m a}\chi_m(r_{0_{j}})
\label{eq:selfenergyp}
\end{equation}
\begin{equation}
A_q(r_{0_{l}},r_{0_{k}})=-t\sum_{n \epsilon
q}\chi_n(r_{0_{l}})e^{\imath k_n a}\chi_n(r_{0_{k}})
\label{eq:selfenergyq}
\end{equation}
\begin{equation}
A=A_p+A_q \label{eq:selfenergytotal}
\end{equation}
Finally for numerical calculation of DF of
var($\nu_{\alpha\alpha}$) and var($\nu_{\alpha\beta}$) for $M\gg
1$ and for higher dimension of the system we calculate the Green
function as:
\begin{equation}
G=[E\hat I-\hat H-\hat A]^{-1} \label{eq:green3}
\end{equation}
To perform numerical calculation of the elements of this Green's
matrix we will use Dyson's equation, as in
\cite{Mackinnon85,Verges99}, propagating strip by strip. This
drastically reduced the computational time, because instead to
inverting an $L^2\times M^2$ matrix, we just have to invert $L$
times $L\times M$ matrices. In this way we build the complete
lattice starting from a single strip, and introducing one by one
the interaction with the next strip. Each time we introduce a new
strip we apply the recursion relations of  Dyson's equation, until
we finally obtain the Green function for the complete lattice.
Once we have the Green's function matrix we calculate
var($\nu_{\alpha\alpha}$) and var($\nu_{\alpha\beta}$) according
to Eqs. (\ref{pdr}) and (\ref{pdt}), and obtain their probability
distributions for random potentials. Over 250000 independent
impurity configurations where averaged for each $N$.

\section{var($\nu_{\alpha\alpha})$ and var($\nu_{\alpha\beta}$) vs $w$}
\label{sec:means}

In this section we are going to study the dependance of the
var($\nu_{\alpha\alpha})$ and var($\nu_{\alpha\beta}$) vs disorder
$w$ and vs the number of mode $M$. In Fig.1 we show the behavior
of var($\nu_{\alpha\alpha})$ and var($\nu_{\alpha\beta}$) as a
function of the disorder $w$. Plot is for a sample of $L=400$ and
$M=4$. The crossover define a critical value of the disorder
$w_c$. In Fig.2 we show the dependence of the critical value $w_c$
with the number of modes $M$ for several samples. As one can see
with increasing the number of modes the crossing point moves to
the left and the $w_c$ decreases. This means that in the weak
localized regime, in analogy with 1D systems the ratio of
localization length $\xi$ to the longitudinal size of the sample
$L$ for given modes $M$ follows, in a good approximation, a law of
the form
\begin{equation}
\frac{\xi}{L}\simeq C(M,w_c,E)
\end{equation}
where $C$ is a constant that depends from the $M$, $w_c$ and
energy. With appropriate choice of an effective length
$L_{eff}=L(a+bM^c)$ (with $a=0.967$, $b=0.035$ and $c=2.33$) we
were able to show that all the curves presented in the Fig.2
collapse into universal curve in Q1D system, supporting the
applicability of the hypothesis of single-parameter
scaling\cite{AALR79} in disordered systems. In Fig.3 we plot this
curve for $w_c$ as a function of $1/L_{eff}$. The different values
of modes are specified inside the figure.
\begin{figure}
\includegraphics[width=8cm]{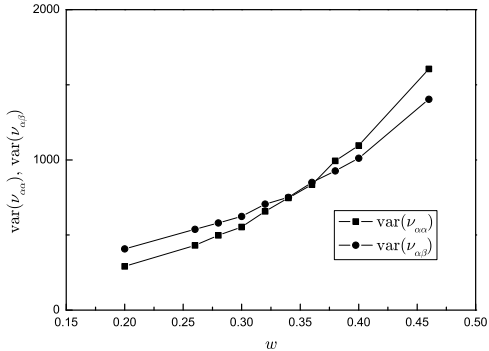}
\caption{Crossover between var($\nu_{\alpha\alpha})$ and
var($\nu_{\alpha\beta}$) for sample of $L=400$ and $M=4$.}
\label{fig1}
\end{figure}
\begin{figure}
\includegraphics[width=8cm]{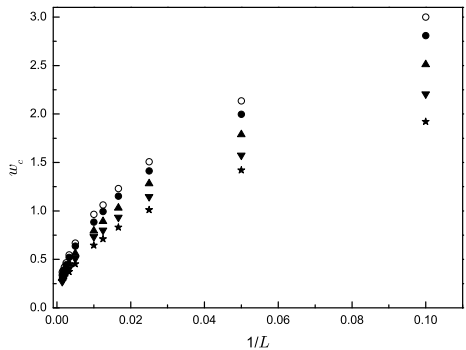}
\caption{Critical value $w_c$  for several samples as a function
of $1/L$: $\circ$ is for 1D sample; $\bullet$, $\blacktriangle$,
$\blacktriangledown$ and $\star$ are for quasi 1D samples with the
numbers of mode $M=2,3,4,5$, respectively.} \label{fig2}
\end{figure}
\begin{figure}
\includegraphics[width=8cm]{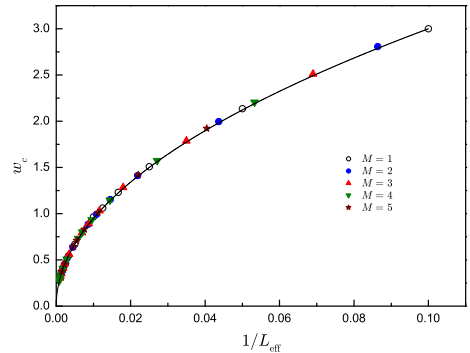}
\caption{Universal curve $w_c$ for several samples as a function
of $1/L_{eff}$. $\circ$ is for 1D sample; $\bullet$,
$\blacktriangle$, $\blacktriangledown$ and $\star$ are for quasi
1D samples with the numbers of mode $M=2,3,4,5$, respectively.}
\label{fig3}
\end{figure}

In strictly 1D system, following \cite{Gasparian96} one can write
($\nu_T\equiv \nu_{\alpha\beta}+\nu_{\beta\alpha}$ and
$\nu_R\equiv \nu_{\alpha\alpha}+\nu_{\beta\beta}$)
\begin{equation}
\left\langle \ln\nu_R\right\rangle =\left\langle
\ln\nu\right\rangle+\left\langle \ln R\right\rangle \label{eq21}
\end{equation}
\begin{equation}
\left\langle \ln\nu_T\right\rangle =\left\langle
\ln\nu\right\rangle+\left\langle \ln T\right\rangle \label{eq22}
\end{equation}
where $R$ and $T$ are the reflection and transmission coefficients
respectively and $\left\langle ...\right\rangle $ denotes
averaging over ensemble. Using the asymptotic behavior of $<\ln
T>$ and $<\ln R>$ as $L\to \infty$ (see, e.g. \cite{Gredeskul})
these expressions in weak disorder regime  can be rewritten as:
\begin{equation}
\left\langle \ln\nu_R\right\rangle =\left\langle
\ln\nu\right\rangle+\ln(1-e^{-2L/\xi}) \label{eq21a}
\end{equation}
\begin{equation}
\left\langle \ln\nu_T\right\rangle =\left\langle
\ln\nu\right\rangle-2L/\xi \label{eq21b}
\end{equation}
In Fig.4 we plot average of $\ln\nu_R$ and $\ln\nu_T$ for
different values of disorder $w_c$ as a function $L/\xi$. We see
that numerical data for these quantities very well coincide with
Eqs. (\ref{eq21a}) and (\ref{eq21b}) for small $w_c$ or for large
$L/\xi$.

\begin{figure}
\includegraphics[width=8cm]{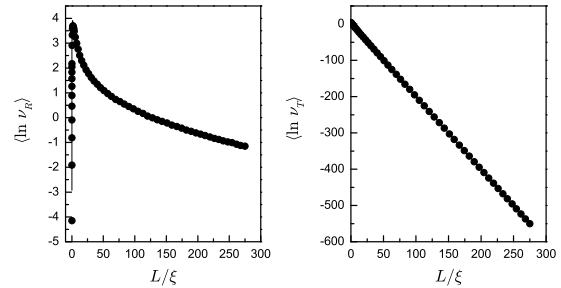}
\caption{Average of $\ln\nu_R$ and $\ln\nu_T$ as a function
$L/\xi$. Solid curves are given by Eqs. (\ref{eq21a}) and
(\ref{eq21b}). The data points ($\bullet$) are the numerical
results for a sample of $L=400$.} \label{fig4}
\end{figure}

\section{Plots and Discussions}
\label{sec:plots}

We are analyzed the DF $P(\nu_{\alpha\alpha})$ and
$P(\nu_{\alpha\beta})$ along the transition from the metallic to
the insulating regime for several samples sizes. We found that the
relative shape of the DF depends only from the disorder parameter
$L/\xi$, i.e. with increasing of the number of modes $M$ we always
can find an appropriate range of $w$ for which all the curves have
the same form. Therefore in the rest of the section, without
losing generality we present our results for a sample of $L=400$
and $M = 4$ for several values of the disorder $w$.

In the metallic regime when the system size much smaller than the
localization length $L<<\xi$ the distribution functions are shown
in Fig.5 with ($W=0.2$, $L/\xi = 0.17$ and $\langle g \rangle =
2.52$). We have checked that the distribution of
$P(\nu_{\alpha\beta})$ is Gaussian-like and can be fit with the
following expression ($B=1.0$, $\mu=116.5$ and $\sigma=20.2$):
\begin{equation}
P(\nu_{\alpha\beta})=
\frac{B}{\sqrt{2\pi}\sigma}e^{-{(\nu_{\alpha\beta}-\mu)^2}/{2\sigma^2}}.
\label{eq:ll}
\end{equation}
In spite of the fact that in our numerical studies we deal with
Q1D systems where the numbers of modes $M>1$, still the
Gaussian-like behavior of the $\nu_{\alpha\beta}$ in ballistic
regime can be understood well if we recall the fact that
$\nu_{\alpha\beta}$ connected with physically meaning full times
characterizing the tunneling process \cite{Gasparian96}. In 1D
systems GPDOS is related to Larmor transmitted time $\tau_T$ (or
Wigner delay time) weighted by the transmission coefficient
\cite{Gasparian96,Prigodin99},
\begin{equation}
\nu_{\alpha\beta}={\frac{T}{2}}\tau_T, \label{nu12}
\end{equation}
The quantity $\tau_T$, which links to the density of states of the
system \cite{GP93} and can be presented
\begin{equation}
\tau_T=\hbar
{Im}\int_0^LG(x,x)\,dx=\hbar{Im}\left\{{\frac{\partial \ln
t}{\partial E}}+{\frac {r+r^{\prime}}{4E}}\right\} \label{total}
\end{equation}
where $G(x,x)$ is the GF for the whole system, $t$ and $r$ are the
transmission and reflection amplitudes from the finite system.
$r^{\prime}$ is the reflection amplitude of the electron from the
whole system, when it falls in from the right.

The second term in Eq.\ (\ref{total}) becomes important for low
energies and/or short systems. This term can be neglected in the
semiclassical WKB case and, of course, when $r$ (and so $r^{\prime
}$) is negligible, e.g., in the resonant case, when the influence
of the boundaries is negligible. Of course the distribution
function of $\nu_{\alpha\beta}$ (Eq. (\ref{nu12})) is affected by
correlations between the value of the DOS (or Wigner delay time)
and the transmission coefficient of resonances via localized
states, but still it can capture some general behavior Wigner
delay time in 1D system in the regime where $T\sim 1$. Wigner
delay time in 1D and in the ballistic regime is given by Gaussian
function and can be characterized by a first moment and a second
cumulant \cite{Texiercomtet99,Heinrichs02}.

Similar relation to Eq. (\ref{nu12}) holds for
$\nu_{\alpha\alpha}$:
\begin{equation}
\nu_{\alpha\alpha}={\frac{R}{2}}\tau_R, \label{nu11}
\end{equation}
where $\tau_R$ characterize the reflection time and defined as:
\begin{eqnarray}
\tau _{R}&=&\hbar {Im}{\frac{1+r}r{e^{-i2\theta(0)}}}
\int_0^LG(x,x)e^{i2\theta (x)}\,dx \nonumber\\
&=& \hbar {Im}\left\{ {\frac{\partial \ln r}{\partial E}}-
{\frac{1-r^2-t^2}{4Er}}\right\} \label{refy1}
\end{eqnarray}
with:
\begin{equation}
\theta (x)= \exp \left (\int_{0}^{x}\frac{dx} {2\:G(x,x)}\right
)\nonumber \label{app1}
\end{equation}
We note that for an arbitrary symmetric potential,
$V((L/2)+x)=V((L/2)-x)$, the total phases accumulated in a
transmission and in a reflection event are the same and so the
characteristic times for transmission and reflection corresponding
to the direction of propagation are equal
\begin{equation}
\tau _{T}=\tau _{R} \label{e:16}
\end{equation}
as it immediately follows from Eqs.\ (\ref{total}) and
(\ref{refy1}). For the special case of a rectangular barrier, Eq.\
(\ref{e:16}) was first found in Ref. \cite{B83}. Comparison of the Eqs.\ (\ref{total}) and (\ref{refy1}%
) shows that for an asymmetric barrier Eq.\ (\ref{e:16}) breaks
down \cite{LA87}.

As one can see from Fig.5 in the same regime DF
$P(\nu_{\alpha\alpha})$ include big range of negative
$\nu_{\alpha\alpha}$ values indicating a predominantly inductive
dynamic response of the system to an external ac electric field
according \cite{BC98}. For positive values of $\nu_{\alpha\alpha}$
the tail of the distribution $P(\nu_{\alpha\alpha})$ is fairly
log-normal with following parameters ($B=0.875$, $\mu=60.5$ and
$\sigma=0.25$):
\begin{equation}
P(\nu_{\alpha\alpha})=\frac{B}{\sqrt{2\pi}\sigma
\nu_{\alpha\alpha}}e^{-{(\ln
\nu_{\alpha\alpha}-\mu)^2}/{2\sigma^2}}. \label{eq:ll2}
\end{equation}

\begin{figure}
\includegraphics[width=8cm]{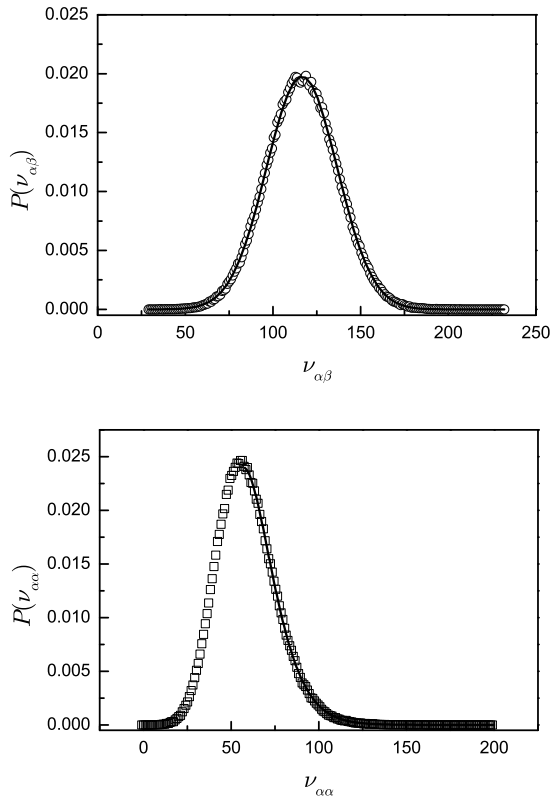}
\caption{Probability distributions of $\nu_{\alpha\alpha}$ and
$\nu_{\alpha\beta}$ in the metallic regime ($\langle g \rangle
=2.52$) for a disorder of $w = 0.2$. The solid lines correspond to
a gaussian distribution for $\nu_{\alpha\beta}$ and a log-normal
tail distribution for $\nu_{\alpha\alpha}$.} \label{fig5}
\end{figure}

With increasing the disorder $w$, when we almost are in crossover
regime we obtain a wide range variety of broad distributions as
shown in Fig. 6 where we plot DF for two values of disorder:
$w=0.5$ ($L/\xi = 0.69$ and $\langle g \rangle =0.75$) in the left
panel and $w=0.6$ ($L/\xi = 0.93$ and $\langle g \rangle =0.5$) in
the right panel. As one can see from Fig.6 (right panel)
$P(\nu_{\alpha\beta})$ has a flat part for almost in all the range
of $\nu_{\alpha\beta}$ while in the left panel it has a strong
decay. In both cases the distributions for $P(\nu_{\alpha\beta})$
can be fitted to two log-normal tails. This type of behavior is
typical also for distribution of conductance $g$ in the same range
of parameters in Q1D, as one can see from the same Fig.6 where we
present $P(g)$. For values $g<1$ we have a plat part while in the
regime $g>1$ we get for distribution of conductance a strong
decay. This is in complete agreement with a number of numerical
simulation in the intermediate regime (see e.g.
\cite{Ruhlanderal01,Ruhlandersoukoulis01,Gopar02}).
\begin{figure}
\includegraphics[width=9cm]{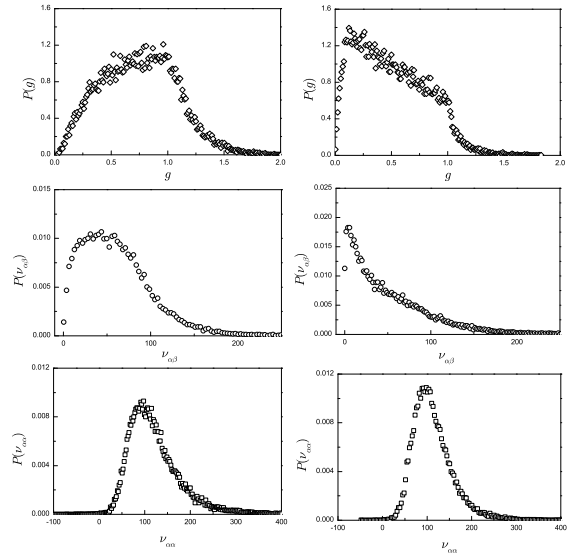}
\caption{Distributions  $P(g)$, $P(\nu_{\alpha\alpha})$ and
$P(\nu_{\alpha\beta})$ in the crossover regime for two value of
the disorder: right panel $w=0.5$ ($L/\xi = 0.69$ and $\langle g
\rangle =0.75$) and left panel $w=0.6$ ($L/\xi = 0.93$ and
$\langle g \rangle =0.5$).} \label{fig6}
\end{figure}

As for $P(\nu_{\alpha\alpha})$ it is shifted to right, to much
larger value of $\nu_{\alpha\alpha}$, which means that it becomes
less conductive. For this range of parameters DF is still quite
symmetric (right panel) but more wider if we compare with the DF
from the Fig.5. The $P(\nu_{\alpha\alpha})$ for $w=0.6$ becomes
less symmetric (left panel).

Further increase of the disorder $w$ (in the insulating region)
$P(\nu_{\alpha\beta})$ becomes a one-side log-normal distribution.
This type of behavior was predicted for distribution of
conductance $g$ in \cite{Muttalib99,Gopar02} and numerically
calculated in \cite{Ruhlanderal01,Ruhlandersoukoulis01,Garcia01}.

With regard to $P(\nu_{\alpha\alpha})$, we can mention that the
tail of the distribution follows a power-law decay
$P(\nu_{\alpha\alpha})\propto 1/\nu_{\alpha\alpha}^m$, with
$m\simeq 2.3$. On the other hand, as $w$ increases
$P(\nu_{\alpha\beta})$ shows a tail in the negative region of
$\nu_{\alpha\beta}$. In Fig.7 we plot the distribution
$P(\nu_{\alpha\alpha})$ and $P(\nu_{\alpha\beta})$ for a disorder
$w=1$ ($L/\xi = 2.6$ and $\langle g \rangle =0.08$).

Deeply in the localized regime ($L \gg \xi$ and $\langle g \rangle
\approx 0$) the distribution of $\nu_{\alpha\beta}$ is log-normal
as one can from Fig.8 where we fit $P(\ln \nu_{\alpha\beta})$ to a
Gaussian distribution:
\begin{equation}
P(x)=\frac{B}{\sqrt{2\pi}\sigma x}e^{-{(\ln
x-\mu)^2}/{2\sigma^2}}, \label{eq:ll3}
\end{equation}
with $B=0.997$, $\mu=-460.5$ and $\sigma=27.7$.

The shape of $P(\nu_{\alpha\alpha})$ is highly asymmetric with two
peaks very closed each other. The position and the high of this
peaks depend on the disorder parameter and cause several shapes of
the distribution function. The tail of the distribution follows a
power-law decay $P(\nu_{\alpha\alpha})\propto
1/\nu_{\alpha\alpha}^m$, with $m\simeq 2.0$

\begin{figure}
\includegraphics[width=8cm]{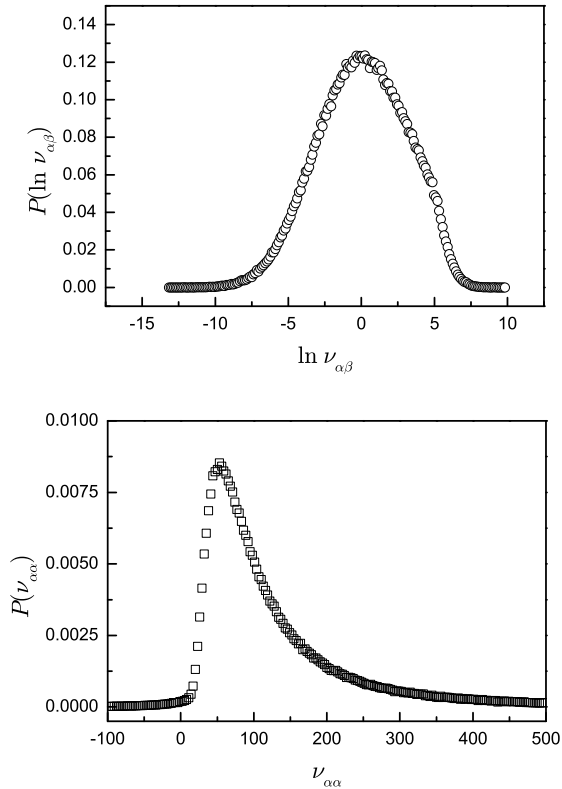}
\caption{Distributions  $P(\nu_{\alpha\alpha})$ and $P(\ln
\nu_{\alpha\beta})$ in the insulating regime ($\langle g \rangle =
0.08$) for a disorder of $W = 1$.  We have a one-side log-normal
distribution for $\nu_{\alpha\beta}$ and power-law tail for
$\nu_{\alpha\alpha}$, $P(\nu_{\alpha\alpha})\propto
1/\nu_{\alpha\alpha}^m$, with $m\simeq 2.3$}. \label{fig7}
\end{figure}

\begin{figure}
\includegraphics[width=8cm]{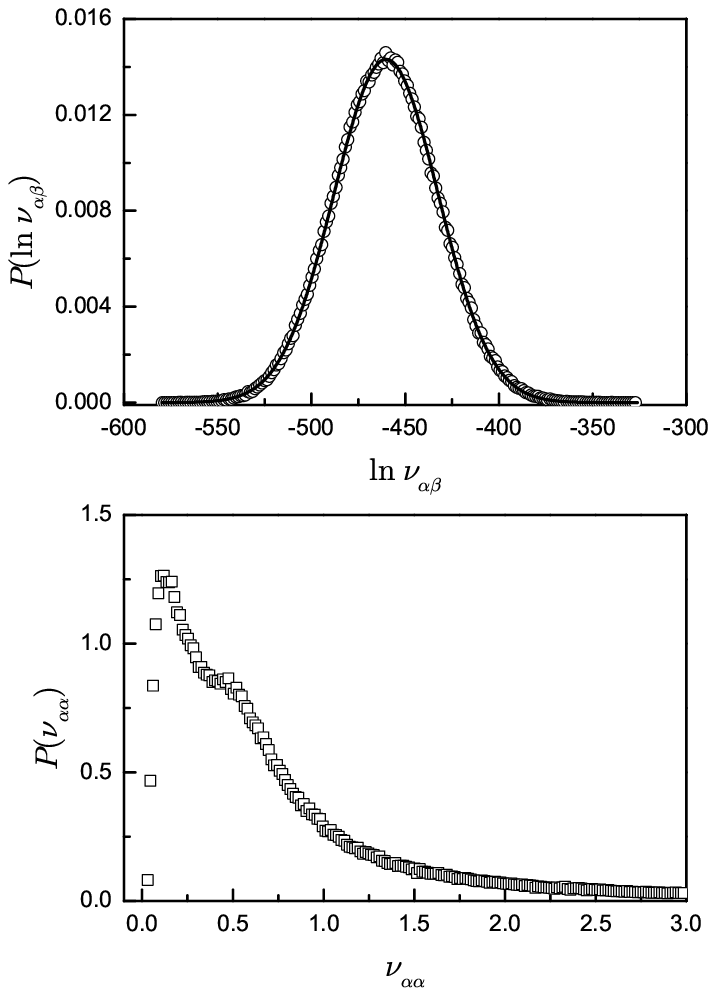}
\caption{Distributions  $P(\nu_{\alpha\alpha})$ and $P(\ln
\nu_{\alpha\beta})$ in the insulating regime ($\langle g \rangle
\approx 0$) for a disorder of $W = 12$. The solid lines correspond
to a gaussian distribution for $\ln \nu_{\alpha\beta}$ that point
out a log-normal distribution for $\nu_{\alpha\beta}$. The tail of
the distribution follows a power-law decay
$P(\nu_{\alpha\alpha})\propto 1/\nu_{\alpha\alpha}^m$, with
$m\simeq 2.0$} \label{fig8}
\end{figure}

\section{Conclusions}
\label{sec:conclusions}

We study the distributions functions for global partial density of
states in quasi-one-dimensional disordered wires as a function of
disorder parameter from metal to insulator. We consider two
different models for disordered Q1D wire: a set of two dimensional
$\delta$ potentials with an arbitrary signs and strengths placed
randomly, and a tight-binding Hamiltonian with several modes and
on-site disorder. It was shown that the poles of Green functions
can be presented as determinant of the rank $N\times N$, where $N$
is the number of scatters. We show that the variances of partial
global partial density of states in the metal to insulator
crossover regime are crossing. The critical value of disorder
$w_c$ where we have crossover can be used for calculation a
localization length in Q1D systems. With increasing the numbers of
mode the crossing point moves to the left and the $w_c$ decreases.

In the metallic regime when the system size much smaller than the
localization length $L<<\xi$ the distribution function for
$P(\nu_{\alpha\beta})$ is Gaussian-like. In the same regime the
distribution function of $P(\nu_{\alpha\alpha})$ is include big
range of negative $\nu_{\alpha\alpha}$ values indicating a
predominantly inductive dynamic response of the system to an
external ac electric field according \cite{BC98}. For positive
values of $\nu_{\alpha\alpha}$ the tail of the distribution
$P(\nu_{\alpha\alpha})$ is fairly log-normal. Almost in crossover
regime the distribution function for $P(\nu_{\alpha\beta})$ can be
fitted to two log-normal tails. As for $P(\nu_{\alpha\alpha})$ it
is shifted to right, to much larger value of $\nu_{\alpha\alpha}$,
which means that it becomes less conductive. Further increase of
the disorder $w$ (in the insulating region) $P(\nu_{\alpha\beta})$
becomes a one-side log-normal distribution. With regard to
$P(\nu_{\alpha\alpha})$, we can mention that the tail of the
distribution follows a power-law decay
$P(\nu_{\alpha\alpha})\propto 1/\nu_{\alpha\alpha}^m$, with
$m\simeq 2.3$. Deeply in the localized regime ($L \gg \xi$ and
$\langle g \rangle \approx 0$) the distribution of
$\nu_{\alpha\beta}$ is log-normal and while the shape of
$P(\nu_{\alpha\alpha})$ is highly asymmetric with two peaks very
closed each other. The position and the high of this peaks depend
on the disorder parameter and cause several shapes of the
distribution function.

\section{Acknowledgments}
\label{sec:acknowledgments}

One of the authors (V.G.) thanks M.~B\"uttiker for useful
discussions and acknowledges the kind hospitality extended to him
at the Murcia and Geneva Universities. J. R. thanks the FEDER and
the Spanish DGI for financial support through Project No.
FIS2004-03117.

\begin{widetext}
\section{Appendix: Dyson equation in Q1D disordered system and the poles of Green's Function}
We consider the Q1D wire with the impurities potential of the
form:
\begin{equation}
V(x,y)=\sum_{n=1}^{N}V_{n}\delta (x-x_{n})\delta (y-y_{n}),
\label{4a}
\end{equation}
where $V_n$, $x_n$ and $y_n$ are arbitrary parameters. The
equation for the Green function with above potential $V(x,y)$ is:
\begin{equation}
\left \{E-\left [-\frac{\hbar^2}{2m}\left
(\frac{d^2}{dx^2}+\frac{d^2}{dy^2}\right )+V_c(y)+V(x,y)\right
]\right\}G(xy;x^{\prime }y^{\prime })= \delta(x-x^{\prime
})\delta(y-y^{\prime }),  \label{1a}
\end{equation}
where the confinement potential $V_c(y)$ depends only on the
transverse direction $y$. The Dyson equation for a Q1D wire can be
written in the form \cite{bagwell90}:
\begin{equation}
G_{ac}(x,x^{\prime })=G_{a}^{0}(x,x^{\prime })\delta
_{ac}+\sum_{b,d}\int G_{a}^{0}(x,x^{\prime \prime })\delta
_{ab}V_{bd}(x^{\prime \prime })G_{dc}(x^{\prime \prime },x^{\prime
})\,dx^{\prime \prime },  \label{1aa}
\end{equation}
The matrix $V_{ab}(x)$ elements of the defect potential are:
\begin{equation}
V_{ab}(x)=\int \chi _{a}^{\ast }(y)V(x,y)\chi
_{b}(y)\,dy=\sum_{n=1}^{N}V_{ab}^{(n)}\delta(x-x_{n}), \label{2a}
\end{equation}
and $V_{ab}^{(n)}$ defined as:
\begin{equation}
V_{ab}(x)=\chi _{a}^{\ast }(y_n)V_n\chi _{b}(y_n)\ \label{3a}
\end{equation}

 Details on the calculation of the GF
$G_{nm}(x,x^{\prime})$ of Dyson equation (\ref{1a}) for this case,
based on the method developed in \cite{GA88,AG91} will be done
elsewhere \cite{RG06}. Here we present main results of calculation
which will be used in numerical calculations. The pole of GF can
be rewritten as a determinant of the rank ($MN\times MN$) ($M$ is
the number of modes and $N$ is the number of delta potentials) The
matrix elements of determinant's $(D_{MN})_{ln}$ are:
\begin{equation}
(D_{MN})_{nl} =
-\mathit{I}\delta_{nl}+(\mathit{I}-\delta_{nl})\{\lambda_{nl}\}\{r^{\left(
l\right) }\} \label{pole}.
\end{equation}
Here:
\begin{equation}
\mathit{I}=\left(
\begin{array}{ccc}
1 &\hdots & 0 \\
\vdots & \ddots &\vdots\\
0& \hdots & 1
\end{array}
\right)  \label{s1}
\end{equation}
is unit matrix. The $l$th scattering matrix $\{r^{\left( l\right)
}\}$ and $\{\lambda_{nl}\}$ matrix are matrices $M\times M$ and
defined in the following way:
\begin{equation}
\{r^{\left( l\right) }\}=\left(
\begin{array}{ccc}
r_{11}^{\left( l\right) } & \cdots & r_{1M}^{\left( l\right) } \\
\vdots & \ddots &\vdots\\
r_{M1}^{\left( l\right) } &\cdots & r_{MM}^{\left( l\right) }
\end{array}
\right),  \label{s1a}
\end{equation}
\begin{equation}
\{\lambda_{nl}\}=\left(
\begin{array}{ccc}
\lambda^{(1)}_{nl} &\hdots & 0 \\
\vdots & \ddots &\vdots\\
0 & \cdots& \lambda^{(M)}_{nl}
\end{array}
\right)=\left(
\begin{array}{ccc}
e^{ik_{1}|x_n-x_l|} &\hdots& 0 \\
\vdots & \ddots &\vdots\\
0 &\cdots &e^{ik_{M}|x_n-x_l|}
\end{array}
\right),  \label{s1b}
\end{equation}
The quantities $r_{mm}^{(l)}$ and $r_{km}^{(l)}$ are:
\begin{equation}
r_{mm}^{(l)}={\frac{V_{mm}^{(l)}G_{m}^{0}(x_{l},x_{l})}{1-
\sum_{n}^{M}V_{nn}^{(l)}G_{n}^{0}(x_{l},x_{l})}}  \label{16a}
\end{equation}
\begin{equation}
r_{km}^{(l)}={\frac{V_{km}^{(l)}\sqrt{
G_{k}^{0}(x_{l},x_{l})G_{m}^{0}(x_{l},x_{l})}}{1-%
\sum_{n}^{M}V_{nn}^{(l)}G_{n}^{0}(x_{l},x_{l})}}  \label{20b}
\end{equation}
respectively. $r_{mm}^{(l)}$ ($m=1,2,...M$) is the complex
amplitude of the reflection of an electron from the isolate
potential $V_{l}$ with the coordinates ${x_{l},y_{l}}$. Electron
incidents from the normal mode $m$ on the left (right) and
reflected normal mode $m$ on the left (right). $r_{km}^{(l)}$ is
the complex reflection amplitude of an electron from the same
$V_{l}$ but it incidents from the normal mode $m$ on the left
(right) side and reflected normal mode $k$ on the same side: By
permuting indexes $k$ and $m$ in (\ref{20b}) one can find the
complex amplitude $r_{mk}^{(l)}$. Note that determinant of the
matrix ${r^{(l)}}$ is zero, i.e.
\begin{equation}
det\{r^{\left( l\right) }\}=0. \label{eq1}
\end{equation}
This is follows from the fact that
$$r_{mm}^{\left( l\right) }r_{kk}^{\left( l\right) }-r_{mk}^{\left( l\right) }r_{km}^{\left( l\right) }=0,$$
which can be checked directly if one used the definitions of
(\ref{20b}) and (\ref{16a}). The rank ($MN\times MN$) of the above
determinant (see (\ref{pole})), after some mathematical
manipulation can be reduced to the determinant of the rank
($N\times N$), as in the case of 1D chain of arbitrary arranged
potentials \cite{GA88,AG91}, with the following matrix elements:
\begin{equation}
(D_{N})_{nl}
=-\delta_{nl}+(1-\delta_{nl})\left[r_{11}^{\left(l\right)}\lambda_{nl}^{\left(1\right)}+{\frac{1}
{r_{11}^{\left(1\right)}}}\sum_{p=2}^M
r_{1p}^{\left(1\right)}r_{p1}^{\left(p\right)}\lambda_{nl}^{\left(p\right)}\right]
\label{deter}.
\end{equation}
Once we know the explicit form of $(D_{N})_{nl}$, we can calculate
the scattering matrix elements without determining the exact
electron wave function in disordered Q1D wire. For example the
transmission amplitude $T_{11}^{(N)}$ from the set of $N$ delta
potentials is:
\begin{equation}
T_{11}^{(N)}=e^{ik_{1}|x_N-x_1|}\frac
{(\bar{D}_{N})_{nl}}{(D_{N})_{nl}}. \label{trans}
\end{equation}
where the matrix $(\bar{D}_{N})_{nl}$ is obtained from the matrix
$({D}_{N})_{nl}$ (Eq. (\ref{deter})) by augmenting it on the left
and on the top in the following way:
\begin{equation}
(\bar{D}_{N})_{nl} = \left| \begin{array}{cccc}
1& r_{11}^{\left( l\right)}&\hdots& r_{11}^{\left( N\right)}e^{ik_{1}|x_N-x_1|} \\
1 & \hdots &\hdots & \hdots \\
\vdots&\vdots &(D_{N})_{nl} & \\
 e^{-ik_{1}|x_N-x_1|} &\vdots & & \end{array} \right|\label{t}.
 \end{equation}
The reflection amplitude $R_{11}^{(N)}$ of electrons from the same
set of $N$ delta potentials is given by:
\begin{equation}
R_{11}^{(N)}=\frac {(\tilde{D}_{N})_{nl}}{(D_{N})_{nl}}.
\label{ref}
\end{equation}
where the matrix $(\tilde{D}_{N})_{nl}$ is obtained from the
matrix $({D}_{N})_{nl}$ (Eq. (\ref{deter})) by augmenting it on
the left and on the top:
\begin{equation}
(\bar{D}_{N})_{nl} = \left| \begin{array}{cccc}
0& r_{11}^{\left( l\right)}&\hdots& r_{11}^{\left( N\right)}e^{ik_{1}|x_N-x_1|} \\
1 & \hdots &\hdots & \hdots \\
\vdots&\vdots &(D_{N})_{nl} & \\
 e^{ik_{1}|x_N-x_1|} &\vdots & & \end{array} \right|\label{t1}.
 \end{equation}
It can checked directly that Eq. (\ref{trans}) for the case of two
point scatterers (i.e. $N=2$) and for two modes ($M=2$) lead us to
($a_1=x_2-x_1$)
\begin{equation*}
T_{11}=e^{ik_{1}a_{1}}
\frac{(1+r_{11}^{(1)})(1+r_{11}^{(2)})+r_{12}^{(1)}r_{21}^{(2)}e^{i(k_{2}-k_{1})a_{1}}-
r_{22}^{(1)}r_{22}^{(2)}e^{2ik_{2}a_{1}}-e^{i(k_{1}+k_{2})a_{1}}
r_{12}^{(2)}r_{21}^{(1)}}{1-r_{11}^{(1)}r_{11}^{(2)}e^{2ik_{1}a_{1}}
-r_{22}^{(1)}r_{22}^{(2)}e^{2ik_{2}a_{1}}-\left(
r_{12}^{(1)}r_{21}^{(2)}+r_{21}^{(1)}r_{12}^{(2)}\right)e^{i(k_{1}+k_{2})a_{1}}}
\end{equation*}
which, after appropriate notation used in \cite{Kumar91}, will
coincides with their expression of $T_{11}$ calculated by transfer
matrix method.

For $R_{11}^{(N)}$ from Eq. (\ref{ref}) we will get
\begin{equation*}
R_{11}=
\frac{r_{11}^{(1)}+r_{11}^{(2)}(1+2r_{11}^{(1)})e^{2ik_{1}a_{1}}
+\left(r_{12}^{(1)}r_{21}^{(2)}+r_{21}^{(1)}r_{12}^{(2)}\right)e^{i(k_{1}+k_{2})a_{1}}}{1-r_{11}^{(1)}r_{11}^{(2)}e^{2ik_{1}a_{1}}
-r_{22}^{(1)}r_{22}^{(2)}e^{2ik_{2}a_{1}}-\left(
r_{12}^{(1)}r_{21}^{(2)}+r_{21}^{(1)}r_{12}^{(2)}\right)e^{i(k_{1}+k_{2})a_{1}}}
\end{equation*}

To close this section let us note that to get the expressions for
the pole of the GF Eq. (\ref{pole}), for transmission amplitude
$T_{11}^{(N)}$ and for $R_{11}^{(N)}$ in tight-binding model one
must to replace the unperturbed GF for normal mode $m$
\begin{equation}
G_m^0(x,x^{\prime})=-i{\frac{m_0}{\hbar^2 k_m}}\exp \left
(ik_a|x-x^{\prime}|\right )
\end{equation}
with
\begin{equation}
k_m=+\sqrt{\frac{2m_0(E-E_m)}{\hbar^2}}
\end{equation}
by the appropriate GF for tight-binding model \cite{Economou83}:
\begin{equation}
G_m^0(l,n)=-\frac{i}{\sqrt{B^2-(E-\epsilon)^2}}e^{|l-n|\Theta}
\end{equation}
Here ($x\equiv (E-\epsilon)/B$)
\begin{equation}
\Theta=\ln (x-i\sqrt{1-x^2})
\end{equation}
and symbol $\sqrt{1-x^2}$ denotes the positive square roots.
\end{widetext}


\begin{thebibliography}{99}
\bibitem{RG06} The details and results will be presented elsewhere.
\end{thebibliography}
\end{document}